\documentclass[reprint,
 amsmath,amssymb,
 aps,prc,
]{revtex4-1}

\usepackage{graphicx}
\usepackage{dcolumn}
\usepackage{bm}
\usepackage{braket}
\usepackage{amsthm}
\usepackage{caption}
\usepackage{array}
\usepackage{color}

\newcommand{\abs}[1] {\ensuremath{ \left| #1 \right| }}

\makeindex
\begin{document}

\preprint{APS/123-QED}

\title{Magnetic quadrupole transitions in the relativistic energy density functional theory}%

\author{G. Kru{\v z}i{\' c}} \email{goran.kruzic@ericsson.com}
 \affiliation{Research Department, Ericsson - Nikola Tesla, Krapinska 45, HR - 10000, Zagreb, Croatia}
\author{T. Oishi}
 \affiliation{Yukawa Institute for Theoretical Physics, Kyoto University, Kyoto 606-8502, Japan}
 \affiliation{Department of Physics, Faculty of Science, University of Zagreb, Bijeni{\v c}ka c. 32, HR-10000, Zagreb, Croatia}
\author{N. Paar}%
 \email{npaar@phy.hr}
 \affiliation{Department of Physics, Faculty of Science, University of Zagreb, Bijeni{\v c}ka c. 32, HR-10000, Zagreb, Croatia}

\begin{abstract}
{\noindent
{\bf Background:} 
Magnetic quadrupole (M2) excitation represents a fundamental feature in atomic nucleus associated to nuclear magnetism induced by spin and orbital transition operator. So far it has only been investigated within the non-relativistic theoretical approaches, and available experimental data are rather limited.
\\
{\bf Purpose:}
We aim to investigate the properties of M2 transitions in closed and open-shell nuclei using the framework of relativistic nuclear energy density functional. The calculated M2 transition strengths could be used to constrain the quenching of the spin gyromagnetic factors.
\\
{\bf Methods:}
The nuclear ground state is calculated with relativistic Hartree-Bogoliubov model, while the M2 excitations are described 
using the relativistic quasiparticle random phase approximation (RQRPA) with the residual 
interaction extended with the isovector-pseudovector term.
\\
{\bf Results:}
The M2 transition strength distributions are described and analyzed for closed shell 
nuclei $^{16}$O, $^{48}$Ca, $^{208}$Pb, open-shell $^{18} \rm O$, $^{42}$Ca, $^{56}$Fe, and 
semi-magic $^{90}$Zr. Detailed analysis of pronounced peaks for magic
nuclei provides the insight into their collectivity. The results are compared with available 
experimental data and the strength missing from the experiment is discussed. 
The evolution of M2 transition properties has been investigated within the $^{36-64} \rm Ca$ 
isotope chain.
\\
{\bf Conclusion:} 
The main M2 transitions have rather rich underlying structure and their
collectivity increases with the mass number due to larger number of 
contributing particle-hole configurations.
Pairing correlations in open shell nuclei have strong effect, causing 
the M2 strength reduction and shifting of the centroid energies to higher values.
The analysis of M2 transition strengths indicate that considerable amount of experimental
strength may be missing, mainly due to limitations to rather restricted energy ranges. The calculated M2 strengths for Ca isotopes, together with the future
experimental data will allow constraining the quenching of the $g$ factors in nuclear medium.
}
\end{abstract}

\pacs{Valid PACS appear here}
\maketitle
\section{Introduction} \label{sec:INTRO}
Magnetic transitions probe the spin and isospin degrees of freedom
of atomic nucleus and provide valuable information for a variety of
nuclear properties. A major interest in nuclear magnetic transitions is strongly focused on dipole (M1) excitations which was subject of a variety of previous studies: for details, see Refs. \cite{2010Heyde_M1_Rev, 1995Richter, 2011Fujita_M1_GT, 2017Schwegner_M1} and references therein.
However, the knowledge on higher multipole magnetic transitions, especially magnetic quadrupole (M2) transitions and respective giant-quadrupole resonances (GQR) is rather limited, both from theoretical and experimental side.
Several experimental studies \cite{Dogotar01, Eramzhyan01, Kuchler01, Kokame01, Stroetzel01, Stroetzel02, Cosel01, Lindgren01, Frey01, Peterson01, Boridy01, Friedrich01} have elaborated highly fragmented M2 structure whose strength $\sum B_{M2}(E)$ 
is strongly suppressed compared to the theoretical results obtained either in shell model or random-phase approximation (RPA) \cite{Eramzhyan01, Cosel01, Knuepfer01, Knuepfer02, Ponomarev01, Castel01, Dehesa01, Ring01, Ring02, Krewald01}.

Theoretical models which investigate excitations based on nuclear fluid-dynamics \cite{Holzwarth01, Holzwarth02, Holzwarth03, Holzwarth04} predicted activation of the twist mode, attributed to the orbital transitions caused by an effective rotation operator around $z$-axis, that has also been experimentally studied in Ref. \cite{Schwesinger01}. In this mode, nucleon orbitals 
as different fluid layers rotate in opposite directions, 
namely for $z > 0$ layers rotate counterclockwise while for $z < 0$ they rotate clockwise.
Consistency check and comparison between different theoretical models can be done using sum-rule approaches,
e.g., those developed in Refs. \cite{Traini01, Suzuki01, Kurath01, Lipparini01, Nishizaki01}. 
The M2 giant resonances have also been studied 
in early study both theoretically and experimentally in $^{38}$Si, $^{90}$Zr, and $^{208}$Pb, 
using simplified particle-hole model, indicating mass-dependent quenching of the spin
gyromagnetic factor $g_s$ \cite{Knupfer1978}.
Mass-dependent quenching of magnetic transitions in nuclei
has also been addressed in the model where spin-isospin core polarization is treated using 
dimesic function techniques \cite{Toki1980}.
It has been shown that the quenching effect increases with 
increasing mass number, in agreement with the experimental findings \cite{Toki1980}.
In high resolution electron scattering the isovector M2 transitions have been studied in $^{42,44}$Ca \cite{Ranga1984}.
In Ref. \cite{Vdovin1985} quasiparticle phonon model has been employed in studies of M2 transition probabilities in odd-A Sn isotopes.
Apart of M2 excitations, higher multipoles $\lambda > 2$ \cite{Speth}, even going toward M8 \cite{Geesamani01} or
higher spin states like M12 and M14 have been experimentally studied \cite{Lallena01, Bacher01}.
In several theoretical studies with the non-relativistic methods \cite{Cosel01},
a wide fragmentation of the M2 mode can be described only by considering the two-particle-two-hole (2p2h) components.

The aim of this work is to provide the first relativistic-microscopic analysis of nuclear M2 transitions
in the energy density functional (EDF) framework, and explore their relevance for possible
constraining the quenching of the spin gyromagnetic factor.
We have focused on M2 electromagnetic excitations composed from 1p-1h transitions, 
from $0^{+}$ ground state (GS) to $2^{-}$ excited state of even-even nuclei, based on
the formalism of the relativistic EDF theory. In this approach, the ground state is
described with the Relativistic Hartree-Bogoliubov (RHB) model \cite{Niksic01}, 
and M2 transitions are investigated within the relativistic quasiparticle
random phase approximation (RQRPA) recently extended for studies of magnetic 
transitions \cite{Kruzic01, Kruzic02}.
The overview of the RHB model and the RQRPA is elaborated in Sec. \ref{sec:FORMLSM}. 
A general expression for M$\lambda$ transitions and other observables are given in
Sec. \ref{sec:ISOVS}.
The results for quadrupole magnetic transitions in specified nuclear
systems $^{16}$O, $^{48}$Ca and $^{208}$Pb, as well as Ca isotope chain are presented in Sec. \ref{sec:M2TRANS}.
Section \ref{sec:PAIRING} is dedicated to pairing effects on M2 
transitions in open-shell 
nuclei $^{18} \rm O$, $^{42} \rm Ca$ , $^{56} \rm Fe$  and $^{90} \rm Zr$. 
The summary and conclusions are given in Sec. \ref{sec:SUM}.

\section{Formalism} \label{sec:FORMLSM}
Theory framework employed in this work follows the formalism based on 
relativistic energy density functional recently developed in studies
of M1 transitions \cite{Kruzic01, Kruzic02}.
The nuclear ground state represents the basis for studies of excitations, and it is described
by employing self-consistent RHB model as given in Ref. \cite{Niksic01},    
with point-coupling interactions.
The corresponding parameterizations
employed in this study are DD-PC1 \cite{Niksic03} and more recent 
parameterization DD-PCX \cite{Yuksel2019} constrained not only by the 
nuclear ground state properties, but also by using the isoscalar giant 
monopole resonance excitation energy and dipole polarizability in $^{208}$Pb.
  
To describe M2 transitions in nuclei, we employ the RQRPA based on the
point coupling interaction, that contains an additional relativistic
isovector-pseudovector (IV-PV) contact type of residual interaction to account
for the unnatural parity transitions \cite{Kruzic01}, given as an effective Lagrangian density,
\begin {equation}
\begin{aligned}
\mathcal{L}_{\rm IV-PV} = -\frac{1}{2}\alpha_{\rm IV-PV} \lbrack  \bar{\Psi}_{N} \gamma^{5} \gamma^{\mu} \vec{\tau}  \Psi_{N} \rbrack   \lbrack  \bar{\Psi}_{N} \gamma^{5}  \gamma_{\mu} \vec{\tau} \Psi_{N}  \rbrack.
\end{aligned}
\end {equation}
For the DD-PC1 interaction, the IV-PV coupling constant has previously
been constrained to {$\alpha_{\rm IV-PV} = 0.53\cdot 197 =104.41~{\rm MeV fm^3}$}
by minimizing relative error  $\rm \Delta\lesssim 1~MeV $ between experimental M1 
excitation peak and theoretical centroid energies for magic nuclei
$\rm ^{48}Ca$ and $\rm ^{208}Pb$ \cite{Kruzic01}.
The same procedure is performed in this study for the DD-PCX interaction,
resulting with 
{$\alpha_{\rm IV-PV} = 0.63\cdot 197= 124.11~{\rm MeVfm^3}$}.
In this way all the model parameters of the relativistic effective Lagrangian density are fixed for M2 transitions.

For open shell nuclei, pairing correlations are described by using phenomenological Gogny interaction \cite{Berger} implemented as in Ref. \cite{Paar01}, 
\begin{equation}
\begin{aligned}
V^{pp}(1,2) = \sum_{i=1, 2} e^{\lbrack (\vec{r}_{1} - \vec{r}_{2})/\mu_{i}\rbrack^{2}}( W_{i} + B_{i}\hat{P}^{\sigma} - H_{i}\hat{P}^{\tau}\\
                          - M_{i}\hat{P}^{\sigma}\hat{P}^{\tau} ),
\end{aligned}
\end{equation}
with parameters $\mu_{i}$, $W_{i}$, $B_{i}$, $H_{i}$ and $M_{i}$ (i = 1,2) 
from the D1S set \cite{Berger}.
The $V^{pp}$ particle - particle correlations in the RQRPA residual interaction 
has the same phenomenological Gogny form and parameterization as in the ground state calculation using the RHB model.

\subsection{Magnetic transitions}\label{sec:ISOVS}
In the general (Q)RPA framework, the $\lambda$-mode strength $B(\lambda,\omega_{i})$ is represented by a discrete spectrum,
{where $\hbar \omega_{i}$ indicates the $i$th eigenenergy.
The discrete strength $B(\lambda,\omega_{i})$ of operator $\hat{q}_{\lambda \nu}$ ($-\lambda \le \nu \le \lambda$) for spherical systems is calculated as follows \cite{Paar01}:
\begin{equation}
\begin{aligned}
B(\lambda,\omega_{i}) = &\Big\vert \sum_{j_{k} j_{k'}} \Big( X^{i,\lambda 0}_{j_{k} j_{k'}} \langle j_{k}||\hat{q}_{\lambda} || j_{k'} \rangle
\\&
+
(-1)^{j_{k} - j_{k'} + \lambda}Y^{i,\lambda 0}_{j_{k} j_{k'}} \langle j_{k'} ||\hat{q}_{\lambda} || j_{k} \rangle
      \Big) \\
&\times \Big(u_{j_{k}}v_{j_{k'}} + (-1)^{\lambda}v_{j_{k}}u_{j_{k'}} \Big)
\Big\vert^{2},
\end{aligned}
\label{transtrength}
\end{equation}
where $j_{k}$ and $j_{k'}$ are quantum numbers of angular momentum for single-particle (SP) states in the canonical basis.
The $u_{j_{k}}$ and $v_{j_{k}}$ are the RHB occupation coefficients, whereas $X^{i,\lambda 0}_{j_{k} j_{k'}}$ and $Y^{i,\lambda 0}_{j_{k} j_{k'}}$ are the corresponding (Q)RPA amplitudes \cite{Paar01}.
}

The M$\lambda$ operator in the relativistic formalism, which acts on Hilbert space with mixed spin-isospin basis, is given in a block diagonal form \cite{Kruzic01}.
That is,
\begin{equation}
\begin{aligned}
\hat{\mu}_{\lambda\nu} = 
\sum_{k = 1}^{A}
\begin{pmatrix} 
 \hat{\mu}_{\lambda\nu}^{(IS)}(11)_{k}   & 0 \\
 0 & \hat{\mu}_{\lambda\nu}^{(IS)}(22)_{k}  
\end{pmatrix}
\otimes 1_{\tau}(k) \\
 -  \sum_{k = 1}^{A} 
\begin{pmatrix} 
 \hat{\mu}_{\lambda\nu}^{(IV)}(11)_{k}  & 0 \\
 0 & \hat{\mu}_{\lambda\nu}^{(IV)}(22)_{k}
\end{pmatrix}  
\otimes \hat{\tau}_{3}(k),
\end{aligned}
\end{equation}
where $ 1_{\tau}$ and  $\hat{\tau}_{3}$ are the unit and Pauli matrices in the isospin space.
The $\hat{\mu}^{(IS)}_{\lambda \nu}$ and $\hat{\mu}^{(IV)}_{\lambda \nu}$ correspond to the isoscalar (IS) and isovector (IV) part of the $M\lambda$ operator for $k$th nucleon \cite{Suhonen01}.
That is,
\begin{equation}
\begin{aligned}
&\hat{\mu}^{(IS,IV)}_{\lambda \nu}(11)_{k} = \hat{\mu}^{(IS,IV)}_{\lambda \nu}(22)_{k}
\\
&= \frac{\mu_{N}}{\hbar}\Big( \frac{2}{\lambda +1}g^{(IS,IV)}_{\ell}\boldsymbol{\hat{\ell}}_{k} + g^{(IS,IV)}_{s}\boldsymbol{\hat{s}}_{k} \Big)    \nabla \big\lbrack r^{\lambda} Y_{\lambda \nu}(\Omega_{k}) \big\rbrack.
\end{aligned}
\end{equation}
The empirical gyromagnetic factors for the bare proton (neutron) are given as $g^{\pi(\nu)}_{\ell} = 1$ ($0$), and $g^{\pi(\nu)}_{s} = 5.586$ ($-3.826$) \cite{Fujita},
expressed in the nuclear-magneton unit $\mu_{N} =e \hbar/2m_{N}$.
Since calculations are performed in the mixed spin-isospin basis,
the isoscalar and isovector gyromagnetic factors are given
separately for the orbital and spin components.
Namely,
\begin{equation}
\begin{aligned}
g^{IS}_{\ell} &= \frac{g^{\pi}_{\ell} + g^{\nu}_{\ell}}{2} = 0.5, 
&
g^{IS}_{s} &= \frac{g^{\pi}_{s} + g^{\nu}_{s}}{2} = 0.880,
\end{aligned}
\end{equation}
for the IS mode, whereas
\begin{equation}
\begin{aligned}
g^{IV}_{\ell} &= \frac{g^{\pi}_{\ell} - g^{\nu}_{\ell}}{2} = 0.5,
& 
g^{IV}_{s} &= \frac{g^{\pi}_{s} - g^{\nu}_{s}}{2} = 4.706,
\end{aligned}
\end{equation}
for the IV mode.

{The M2 transition strength from the R(Q)RPA is given as
\begin{equation}
B_{M2}(E_{i}) = \abs{\beta^{IS}_{M2}(E_{i}) + \beta^{IV}_{M2}(E_{i})}^{2}.
\end{equation}
Here the IV-M2 amplitude is determined as
\begin{equation}\label{eq:15}
\begin{aligned}
\beta^{IV}_{M2}(E_{i}) = \sum_{j_{k} j_{k'}} \Big( X^{i}_{j_{k} j_{k'}} + (-1)^{j_{k} - j_{k'}}Y^{i}_{j_{k} j_{k'}} \Big) \\
\times \Big(u_{j_{k}}v_{j_{k'}} + v_{j_{k}}u_{j_{k'}} \Big) 
\langle j_{k'} ||\hat{\mu}_{\lambda=2}^{(IV)} || j_{k} \rangle  \tau_{IV},
\end{aligned}
\end{equation}
with $\tau_{IV}=1$ ($-1$) for neutrons (protons).
On the other side, the IS-M2 amplitude reads
\begin{equation}\label{eq:20}
\begin{aligned}
\beta^{(IS)}_{M2}(E_{i}) = \sum_{j_{k} j_{k'}} \Big( X^{i}_{j_{k} j_{k'}} +(-1)^{j_{k} - j_{k'}}Y^{i}_{j_{k} j_{k'}} \Big) \\
\times \Big(u_{j_{k}}v_{j_{k'}} +v_{j_{k}}u_{j_{k'}} \Big) 
\langle j_{k'} || \hat{\mu}_{\lambda=2}^{(IS)} || j_{k}\rangle.
\end{aligned}
\end{equation}
In addition, the spin-M2 strength $B^{\sigma}_{M2}$ and the orbital-M2 strength
$B^{\ell}_{M2}$ are obtained by retaining only the spin and orbital part of the $M2$ transition operator, respectively.
}


{The moments $m_{k}$ of discrete-strength distributions are defined as \cite{Paar01}
\begin{equation} \begin{aligned}
m_{k} = \sum_{i} B_{M2}(E_{i}) E^k_{i}.
\end{aligned} \end{equation}
The non-energy-weighted (NEW) moment, $m_{0}$, corresponds to the total summation}  {of the transition strength,
whereas $m_{1}$ is the energy-weighted (EW) summation of $B_{M2}$ values.} {
The centroid energy is obtained as $\bar{E} = m_{1}/m_{0}$, which represents an average energy of calculated strength distribution.
The (Q)RPA response function $R_{M2}(E)$ is defined as \cite{Paar01},
\begin{equation}\label{eq:5}
\begin{aligned}
R_{M2}(E) = \sum_{i} B_{M2}(E_{i}) \frac{1}{\pi} \frac{\Gamma/2}{ (E - \hbar\omega_{i})^{2} + (\Gamma/2)^{2}},
\end{aligned}
\end{equation}
where the discrete spectrum is smoothed by the Lorentzian distribution using the width $\Gamma = 1.0 \rm \ MeV$.
Note that the width of the individual excitation can not be considered in this framework, and further developments going beyond the (Q)RPA are required \cite{1993Kam,Schwengner01}.
}

 {For the discussion of our results,
we also mention the Weisskopf estimate for the transition strength,
that corresponds to the single-particle (SP) electric or magnetic transition \cite{1951Weisskopf,RingSchuck}.
}
The Weisskopf estimate for the proton's $M\lambda$ strength is
\begin{equation}
  B_{SP-M\lambda} \cong \frac{1+(g^{\pi}_s/2)^2}{\pi} \left(\frac{3}{\lambda+3}  \right)^{2} R_0^{2\lambda-2}\mu^2_{\rm N}
\end{equation}
with $R_0 \equiv 1.2A^{1/3}$ fm.
Note that the first factor is replaced to $(g^{\nu}_s/2)^2/\pi$ for neutrons.
Thus, the SP-M2 strength is roughly proportional to $A^{2/3}$,
where $A$ is the mass number.
That is,
\begin{equation}
  B_{SP-M2} \cong 1.45 A^{2/3}~~\mu^2_{\rm N}{\rm fm}^2.
\end{equation}
This is used as the M2 Weisskopf unit.
For one excited state with the strength $B_{M2}(E)$, the ratio of
$B_{M2}(E)/B_{SP-M2}$ gives a rough estimate of the number of excited nucleons.

\begin{figure}[tb]
\includegraphics[width=0.88\hsize]{M2_O16_IS_IV_MTD4_EPS600.eps}\\
\includegraphics[width=0.88\hsize]{M2_O16_L_S_MTD4_EPS600.eps}
\includegraphics[width=0.9\hsize]{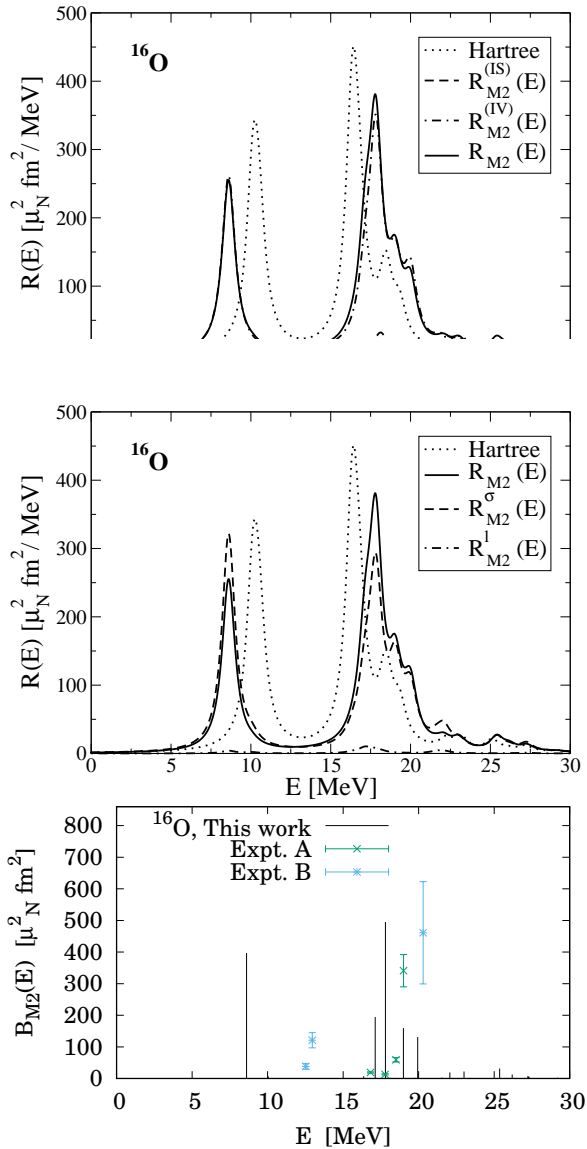}
\caption{(Top) The M2 strength distribution for $^{16}$O based on the RRPA with DD-PC1 interaction: $R_{M2}(E)$, $R^{(IS)}_{M2}(E)$, and $R^{(IV)}_{M2}(E)$ represent the full, IS, and IV-M2 response functions, respectively.
The unperturbed full response at the Hartree level is also shown.
(Middle) The same plot but for the $R_{M2}(E)$ (full), $R^{\sigma}_{M2}(E)$ (spin) and $R^{\ell}_{M2}(E)$ (orbital) response functions for $^{16}$O.
(Bottom) The discrete M2-strength distribution for $^{16}$O.
Experiments A and B indicate Refs. \cite{Kuchler01} and \cite{Stroetzel01, Stroetzel02}, respectively.}
\label{fig:O02}
\end{figure}

\section{Results and discussions}\label{sec:M2TRANS}
In this work the $0^{+}\rightarrow 2^{-}$ transitions induced by the M2 operator have been evaluated for spherical nuclei.
Calculated results are compared either with available experimental data, different theoretical approaches or both.

\begingroup\renewcommand{\arraystretch}{1.5}
\begin{table}[tb]
\begin{center}
\begin{tabular}{lll}
\hline  \hline
  Energy~[MeV]   &$B_{M2}(E)$~[$\mu^{2}_{\rm N}{\rm fm}^{2}$]~~ &Reference  \\
~  12.53   &$38 \pm 9$  &\cite{Stroetzel01, Stroetzel02}  \\
~  12.96     &  $121 \pm 24$  &\cite{Stroetzel01, Stroetzel02}  \\
~  16.82     &  $19 \pm 2$ &\cite{Kuchler01} \\
~  17.78     &  $13 \pm 2$ &\cite{Kuchler01} \\
~  18.50     &  $59 \pm 7$ &\cite{Kuchler01} \\
~  19.0   &$341 \pm 51$ &\cite{Kuchler01} \\
~  20.30     &  $461 \pm 162$ &\cite{Stroetzel01, Stroetzel02} \\
\hline
~    &$\sum B_{M2}$ & \\
~12.53-20.30~~~~~  &$1052 \pm 257$ &\cite{Kuchler01}+\cite{Stroetzel01, Stroetzel02} \\
~0-50     &$1534.58$  &(This work) \\ 
\hline  \hline
\end{tabular}
\end{center}
\caption{Experimental M2-excitation energies and transition strengths for $^{16}$O. The NEW summation of  {of $B_{M2}$ values} from the relativistic RPA with DD-PC1 functional is also presented.
}
\label{table:Exp_O16_01}
\end{table}
\renewcommand{\arraystretch}{1}

\subsection{M2 transitions in $^{16}$O} \label{sec:O16RES}
As the first example, we investigate the doubly magic nucleus $^{16}$O with the RRPA with DD-PC1 interaction \cite{Niksic03}.
Figure \ref{fig:O02} shows the corresponding M2 strength distributions, including $R_{M2}$ (full), $R^{(IS)}_{M2}$, and $R^{(IV)}_{M2}$.
The centroid energy of the full response amounts $\bar{E}= 16.55$ MeV.
The full M2 strength distribution is composed of two main
excitation peaks, namely, low-energy peak at $\approx$ 8 MeV, and higher one at $\approx$ 17.5 MeV.
One can also observe
that the transition strength is dominated by the IV response,
while the IS transitions have small contribution being visible only at high-energy region of the spectra.
{This dominance is mainly due to the spin $g$ factors: $g^{IV}_{s} > g^{IS}_{s}$.
In order to assess the role of the residual RRPA interaction, the unperturbed Hartree response is also shown for comparison with the full calculation.
Clearly, the residual interaction considerably modifies the unperturbed spectra,
resulting in shifts of the energies of the two main peaks.
We also calculate the NEW sum of the transition strength up to 50 MeV.
That reads $m_0 =\sum B_{M2}$ = 1534.58, whereas
$\sum B^{(IV)}_{M2}$ = 1453.83, and 
$\sum B^{(IS)}_{M2}$ = 73.59 in the unit of $\mu^{2}_{N}$ fm$^{2}$.
This again confirms the dominance of IV component.
}

{We decompose the full M2 strength distribution into the spin and orbital components in order to assess the relevance of their contributions.
The corresponding spin and orbital response functions, $R^{\sigma}_{M2}(E)$ and $R^{l}_{M2}(E)$, are shown in Fig. \ref{fig:O02}.
One clearly reads that the spin response dominates over the orbital one through the whole energy range.
However, the interference between those contributions is complicated.
At the low-lying peak at $8.6$ MeV, the spin-M2 response is larger than the total response.
At the higher energies, the opposite result is obtained.
By symbolically writing, the total response is determined as the absolute square of the two transition amplitudes, i.e.,
\begin{equation}
  R_{M2}(E) = \abs{\beta^{\sigma}_{M2}(E) +\beta^{\ell}_{M2}(E)}^2,
\end{equation}
whereas $R^{\sigma /\ell}_{M2}(E) = \abs{\beta^{\sigma /\ell}_{M2}(E)}^2$.
Thus, at the low-lying (high-energy) M2 peak, the spin-M2 and orbital-M2 amplitudes have the opposite (same) phase to realize the reduced (enhanced) total $R_{M2}(E)$ value.
As the result, our NEW summations up to 50 MeV read
$m_0= \sum B_{M2}$ = 1534.6 (full),
$\sum B^{\sigma}_{M2}$ = 1492.2 (spin), and
$\sum B^{\ell}_{M2}$ =  44.3 (orbital) in the unit of $\mu^{2}_{N}$fm$^{2}$.
}

The available experimental data for $^{16}$O from Refs. \cite{Stroetzel01,Stroetzel02,Kuchler01} are listed in Table \ref{table:Exp_O16_01}.
These data indicate a fragmentation of the M2-excited states among 12-20 MeV.
{Note that the M2-Weisskopf unit is $B_{SP-M2} = 9.21~\mu^2_{\rm N}$fm$^2$ in this case \cite{1951Weisskopf,RingSchuck}.
In contrast, the experimental M2 strength shows noticeably larger values than this Weisskopf unit.
Thus, the M2 transition is expected as a collective process, where several SP transitions simultaneously contribute.
These data also suggest that the Weisskopf assumption is not definitely applicable.
For example, at 19.0 MeV with $B^{\rm expt.}_{M2}\cong 341~\mu^2_{\rm N}$fm$^2$ \cite{Kuchler01},
the mean number of excited nucleons is estimated as
$341/9.21 =37$, being larger than the total mass number of $^{16}$O.
Thus, the actual SP-M2 strength is expected to be more dependent on the quantum numbers of relevant orbits.
}
Table \ref{table:Exp_O16_01} also displays the total summation of the experimental data.

 {As shown in Table \ref{table:Exp_O16_01},
the M2-NEW summation of transition strength} 
{from the present RRPA calculation is somewhat larger than the experimental value,
indicating that some strength could be missing from experiments.
Especially, the calculated low-lying M2 state has not been reported in experimental studies.}

In order to analyze the underlying structure of each pronounced 
M2 state obtained in the RRPA calculation,
Table \ref{table:Th_O16_01} in the Appendix is prepared.
There, the major $ph$-transition components contributing to 
the M2 transition strength are listed.
These components, denoted as $b^{\pi,\nu}_{ph}$, correspond to the
contribution of each $ph$ configuration to the summation in the
transition strength, Eq. (\ref{transtrength}), for protons $(\pi)$ and neutrons $(\nu)$, respectively.
Thus, the overall transition strength is given as
\begin{equation}
B_{M2}(E) =  \abs{\sum\limits_{ph} b^{\nu}_{ph}(E) + \sum\limits_{ph} b^{\pi}_{ph}(E)}^{2}.
\end{equation}
The analysis of $b^{\pi,\nu}_{ph}$ provides useful information on
the structure of RPA states and their collective properties, as
shown in previous studies \cite{Paar2009,Roca-Maza2012}. 
In particular, if more $ph$ components have sizable contributions, thus involving excitations
of considerable number of nucleons, one can conclude on collective
nature of the RPA state. On the other side, if e.g., only one
$ph$ component is relevant, the RRPA state is of pure
single-particle nature. By inspecting M2 partial contributions
for $^{16}$O in Table \ref{table:Th_O16_01}, one can conclude that,
all the protons and neutrons in the $1p_{3/2}$ and $1p_{1/2}$ orbits simultaneously contribute, whereas the deepest $1s_{1/2}$ orbit 
is not active. The most collective state is obtained at 17.81 MeV,
with 7 $ph$ configurations contributing to the overall transition
strength. This collectivity is different than for the M1 transitions, where only a
few $LS$-partner orbits can be active \cite{Kruzic01, Kruzic02}. There 
are also other M2 states which are composed
of one or two $ph$ configurations, such as those at
19.04 and 19.99 MeV.

The low-lying M2 peak predicted at $8.6$ MeV is mainly from the
$1p^{-1}_{1/2} \rightarrow 1d_{5/2}$ transitions of both protons and neutrons.
For the absence of this peak in experiments, from the theory side 
one possible reason is the ambiguity of IV-PV residual interaction in QRPA.
The M2-excitation energy is sensitive to the IV-PV coupling, where the 
finite ambiguities remain as discussed in our previous study on M1 and
Gamow-Teller transitions \cite{Oishi2022}.

\begin{figure}[tb]
\includegraphics[width=0.88\hsize]{M2_Ca48_IS_IV_MTD4_EPS600.eps}\\
\includegraphics[width=0.88\hsize]{M2_Ca48_L_S_MTD4_EPS600.eps}
\includegraphics[width=0.9\hsize]{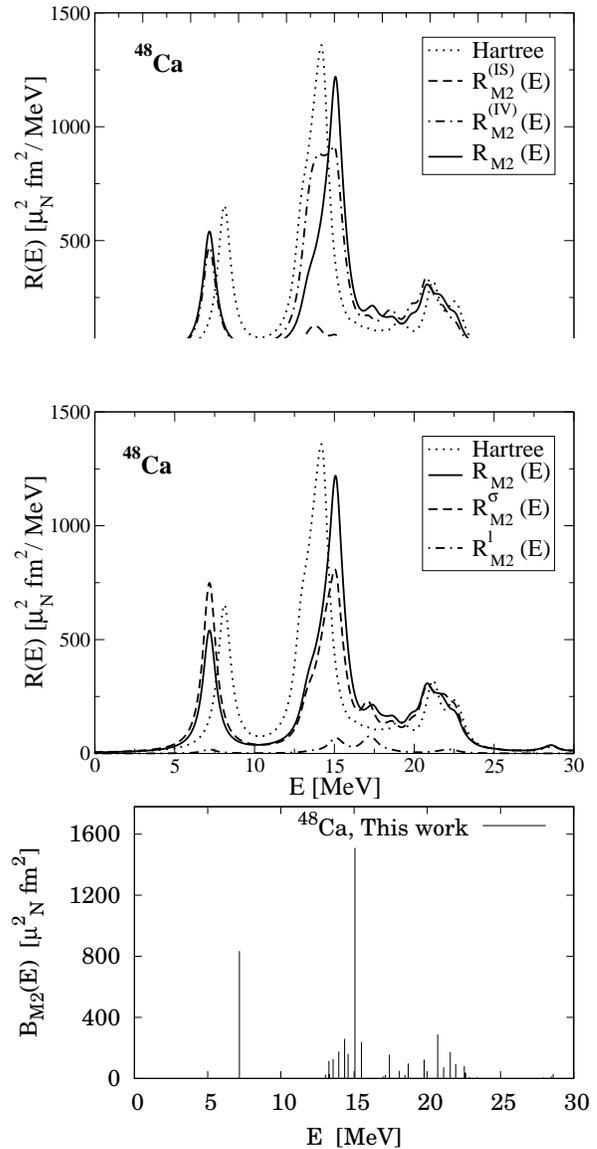}
\caption{The same as Fig.~\ref{fig:O02} but for $^{48}$Ca.}
\label{fig:Ca01}
\end{figure}

\subsection{M2 transitions in $^{48}$Ca} \label{sec:Ca48RES}
The M2-response functions calculated with the RRPA with DD-PC1 interaction for $^{48}$Ca are shown in Fig. \ref{fig:Ca01}. 
The transition strength is characterized by the main peak at excitation energy $\approx$ 15 MeV, and two
additional peaks, i.e., low-lying one at $\approx$ 7 MeV and high-lying
one at $\approx$ 21 MeV.
The calculated centroid energy of the full response amounts $\bar{E}= 15.48$ MeV.
The corresponding NEW summations up to 50 MeV read
$m_0=\sum B_{M2}$ = 4915.5 (full),
$\sum B^{(IV)}_{M2}$ = 5136.4, and 
$\sum B^{(IS)}_{M2}$ = 494.2 in the unit of $\mu^{2}_{N}$ fm$^{2}$.
Figure \ref{fig:Ca01} also shows the full, spin and orbital M2 response functions for $^{48}$Ca.
The corresponding summations give
$\sum B^{\sigma}_{M2}$ = 4575.6, and
$\sum B^{\ell}_{M2}$ =  319.1 in the unit of $\mu^{2}_{N}$ fm$^{2}$.
It shows that the spin-M2 strength is significant, as similarly confirmed in the previous $^{16}$O case.
In addition, one can find that,
at the low-lying (high-energy) M2 peak, the spin-M2 and orbital-M2 amplitudes have the opposite (same) phase to realize the reduced (enhanced) total $R_{M2}(E)$ value.
This conclusion is consistent to the $^{16}$O case.

The calculated EW moment amounts $m_{1} = 76.1\times 10^{3}$ MeV$\mu^{2}_{N}$fm$^{2}$ up to 50 MeV.
This is larger than the RPA result in Ref. \cite{Cosel01}, $52.4\times 10^{3}$ MeV$\mu^{2}_{N}$fm$^{2}$.
Note that the result in Ref. \cite{Cosel01} includes the quenching effect, $g_s \longrightarrow 0.64 g_s$, in order to reproduce the experimental data of M1 transitions \cite{Kruzic02}.
For comparison, the experimental EW-M2 summation amounts $15.7\times 10^{3}$ MeV$\mu^{2}_{N}$fm$^{2}$ up to 15 MeV \cite{Cosel01}.
For the observed discrepancy between theoretical and experimental values in the present study,
one explanation is the quenching effect on $g$ factors.
For adjusting our EW-M2 summation to the experimental one, the quenching factor of
$\zeta=\sqrt{15.7/76.1} \cong 0.45$ is necessary for all the $g$ factors.
However, this strong quenching is not consistent to the M1 result in our previous work \cite{Kruzic01},
where the minor quenching of $\zeta=0.8-0.9$ was appropriate to reproduce the experimental M1 summation.
Another possible reason is the effect going beyond the RPA.
Indeed in Ref. \cite{Cosel01}, the second RPA calculation for M2 transitions was also performed, where the better agreement with experiment than the pure RPA calculation was confirmed.
This also suggests that beyond-RPA effects may be more (less) essential for the M2 (M1) transition, in which the collectivity is relatively strong (weak).
However, similar beyond-RPA calculation but of relativistic version is still challenging, and going beyond the scope of this work.
Note also that considerable M2 data especially of high energies could be missing from the experimental side due to the limited energy range \cite{Cosel01}.

\begin{figure}[t]
\includegraphics[width=\hsize]{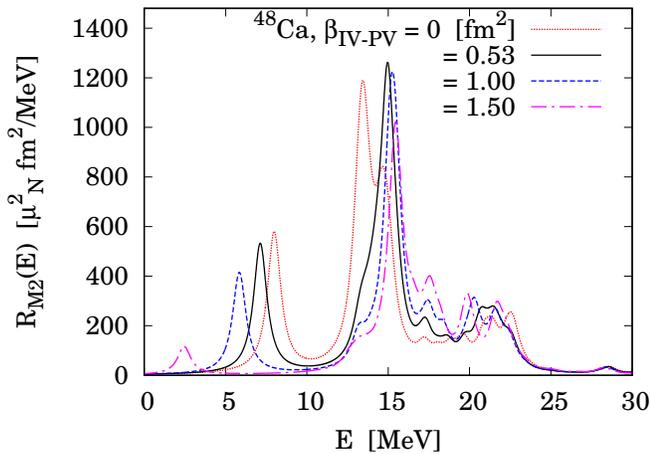}\\
\caption{{The M2 transition strength distributions for $^{48}$Ca by 
changing the IV-PV coupling strength parameter, ${\alpha}_{\rm IV-TV}={\beta}_{\rm IV-PV}\cdot \hbar c$.
Note that ${\beta}_{\rm IV-TV}=0.53$ fm$^2$ is the default setting when
combined with the DD-PC1 interaction.}
}
\label{fig:TPV_check}
\end{figure}

Details on the composition of M2 transitions in $^{48}$Ca are given in Table \ref{table:Th_Ca48_01} in the Appendix.
The analysis of the relevant $ph$ contributions to the main 10 excited states indicates that,
both for protons and neutrons,
the $2s_{1/2}$, $1d_{5/2}$, and $1d_{3/2}$ orbits are active for M2 transitions.
In addition, the neutrons in $\nu 1f_{7/2}$ are also contributing.
On the other side, several transitions, e.g. $1p_{*/2} \rightarrow 1d_{*/2}$, are forbidden,
since their final states are occupied.
Notice that the collectivity of M2 transition is shown as larger for $^{48}$Ca
than previously analyzed $^{16}$O, simply because, in $^{48}$Ca,
the initial state has more orbits of the M2-active nucleons.
In correspondence, the NEW summation $m_0$ amounts the larger value for $^{48}$Ca:
$m_0(^{48}{\rm Ca})/m_0(^{16}{\rm O}) =4915.46/1534.58 \cong 3.2$.

From Table \ref{table:Th_Ca48_01} in the Appendix,
the low-lying peak at 7.17 MeV is predicted with $B_{M2}=834.41~\mu^{2}_{N}$fm$^{2}$.
There, the most dominant contribution is from
the proton transition $1d^{-1}_{3/2} \longrightarrow 1f_{7/2}$.
This transition is natural for the M2-selection rule.
In comparison, however, the corresponding data in Ref. \cite{Cosel01}
between 6-8 MeV yields only $\cong 25~\mu^{2}_{N}$fm$^{2}$.
The RPA and second RPA calculations in Ref. \cite{Cosel01} commonly predict the similar, minor strengths in this low-lying region of $^{48}$Ca.
Thus, the discrepancy in M2 transition strengths
does not seem to be cured even if the beyond-QRPA effects are taken into account.
Alternative reason is possibly the sensitivity of the RRPA energies on the IV-PV part 
of the residual interaction.
By assuming that our RRPA energy, 7.17 MeV, is over-estimated,
the low-lying M2 peak may exist but below the minimum of experimental accessibility.
In Fig. \ref{fig:TPV_check},
the M2 response as a function of the strength parameter of the
IV-PV coupling is presented.
There, one can find that the low-lying M2 peak is sensitive to the
IV-PV interaction, similar to the sensitivity that was also obtained
in the study of M1 and Gamow-Teller transitions \cite{Oishi2022}.
Further experimental studies of magnetic transitions could provide
additional constraints for the IV-PV interaction. From the theory side,
improvement of the relevant SP energies could also result in modifications
of the present results. 

\begin{figure}[tb]
\includegraphics[width=\hsize]{M2_Pb208_IS_IV_MTD4_EPS600.eps}\\
\includegraphics[width=\hsize]{M2_Pb208_L_S_MTD4_EPS600.eps}
\includegraphics[width=\hsize]{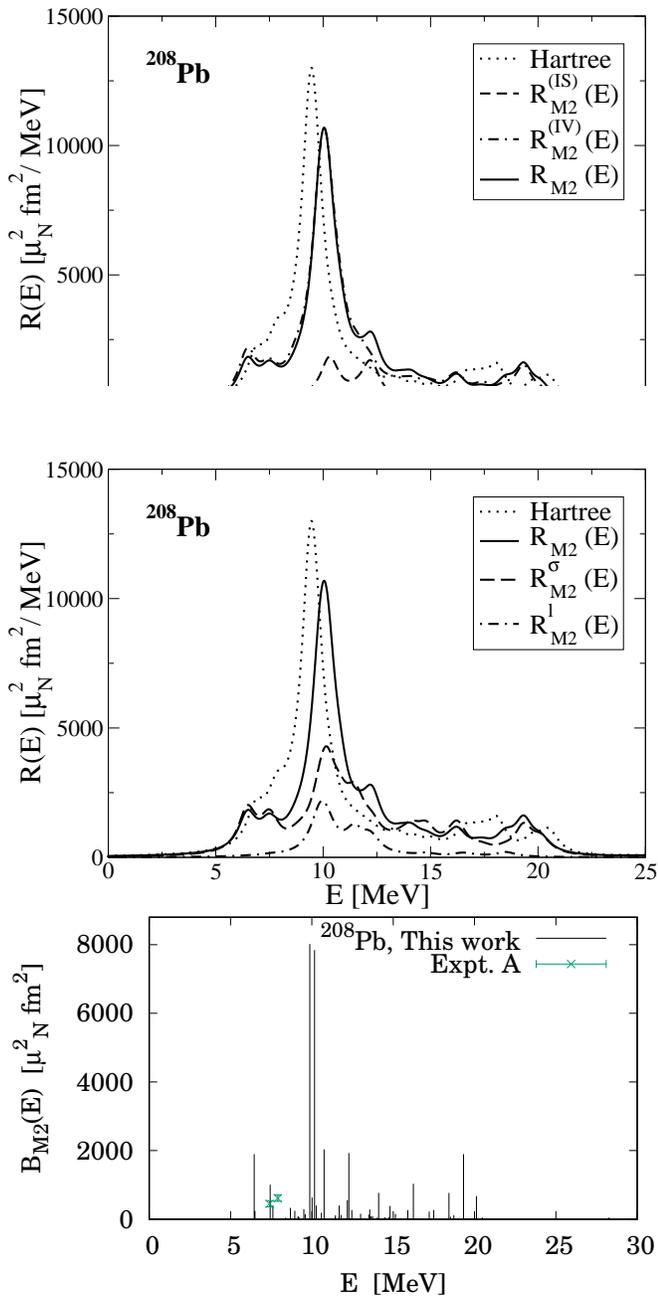}
\caption{The same as Fig.~\ref{fig:O02} but for $^{208}$Pb.
Experiment A indicates Ref \cite{Lindgren01}.}
\label{fig:Pb01}
\end{figure}

\subsection{M2 transitions in $^{208}$Pb}\label{sec:Pb208RES}
Figure \ref{fig:Pb01} shows the full, IS and IV M2-response functions for the $^{208}$Pb nucleus, where one dominant peak and simultaneously several finite structures are confirmed.
The centroid energy of full response amounts 
$\bar{E}^{th.}= 11.87$ MeV, that is slightly above the main peak excitation energy due to the tail of the transition strength extending toward 20 MeV.
The NEW summations up to 50 MeV yield $m_0 =\sum B_{M2}$= 35648 $\mu^{2}_{N}$fm$^{2}$ (full),
$\sum B^{(IV)}_{M2}$ = 35459 $\mu^{2}_{N}$fm$^{2}$, and
$\sum B^{(IS)}_{M2}$ = 6358 $\mu^{2}_{N}$fm$^{2}$.
Figure \ref{fig:Pb01} also displays the decomposition of the full M2 strength into the spin and orbital components. 
In addition to the dominant spin response, a non-negligible contribution from the orbital response is observed, that is, the peaked around 10 MeV.
There, a constructive interference between the spin and orbital components exists as $R_{\rm M2} > R^{\sigma /\ell}_{\rm M2}$.
On the other hand, the destructive interference is also found around 7 MeV, but its effect is minor in this $^{208}$Pb nucleus.
The corresponding summations up to 50 MeV are
$\sum B_{M2} =$  35648 $\mu^{2}_{N}$fm$^{2}$ (full),
$\sum B^{\sigma}_{M2}$ = 25515 $\mu^{2}_{N}$fm$^{2}$ (spin), and
$\sum B^{\ell}_{M2}$ =  6936 $\mu^{2}_{N}$fm$^{2}$ (orbital).

\begingroup\renewcommand{\arraystretch}{1.5}
\begin{table}[tb]
\begin{center}
\begin{tabular}{lll}
\hline
\hline
~Energy  &$B_{M2}(E)$  &Reference  \\
~[MeV]   &[$\mu^{2}_{N}$fm$^{2}$] & \\
~7.40         &  449$\pm$89  &\cite{Lindgren01} \\
~7.91         &  614$\pm$92  &\cite{Lindgren01} \\
\hline
~&$\sum B_{M2}$  & \\
~6.428-8.008~~~~  & 8500 $\pm$ 750  &\cite{Frey01} \\
~7.457-8.008~~~~  & 5230 $\pm$ 130  &\cite{Frey01} \\
~0-50    &35648     &(This work) \\
~0-9.88  &11785     &(This work) \\
~0-8.97  &~3694     &(This work) \\
\hline
\hline
\end{tabular}%
\end{center}
\caption{Experimental M2 data for $^{208}$Pb.
Note that the M2-Weisskopf unit is $B_{SP-M2} = 50.9~\mu^2_{\rm N}$fm$^2$.}
\label{table:Exp_Pb208_02}
\end{table}
\renewcommand{\arraystretch}{1}

In Tables \ref{table:Th_Pb208_01} and \ref{table:Th_Pb208_02} in the Appendix,
particle-hole transition components of the main M2 peaks for $^{208}$Pb are presented.
In comparison to other nuclei previously discussed,
for $^{208}$Pb more complicated composition is obtained,
especially for the two main peaks at 9.88 and 10.18 MeV.
This is expected due to larger number of M2-active nucleons.

The experimental data on M2 excitations in $^{208}$Pb are summarized in Table \ref{table:Exp_Pb208_02}.
In Ref. \cite{Frey01},
if the experimental M2 strengths between 6-8 MeV are summed,
the result reads
$\sum B_{M2}= 8500 \pm 750~\mu^{2}_{N} {\rm fm}^{2}$ \cite{Frey01}.
Note that, in Ref. \cite{Frey01}, the analyzing procedure with RPA refers to the
experimentally observed $J^{\pi} = 2^{-}$ peak at 7.47 MeV,
and its result is confirmed by the MSI-RPA in Ref. \cite{Knuepfer02}.
The other RPA study in Ref. \cite{Ponomarev01} gives $9700$-$12600~\mu^{2}_{N}{\rm fm}^{2}$
within the excitation energies of 6.1-8.4 MeV.
Other theoretical investigations have concluded
11000 $\mu^{2}_{N} {\rm fm}^{2}$ \cite{Castel01} or
11600 $\mu^{2}_{N} {\rm fm}^{2}$ \cite{Dehesa01}.
Indeed the M2 summation depends on the energy window available in measurements.
By considering the present results with the DD-PC1 interaction outlined in our analysis in tables \ref{table:Th_Pb208_01} and \ref{table:Th_Pb208_02},
if we sum up $B_{M2}$ values at excitation energies of $E_{x}=$6.46, 7.45, 7.61, 8.69, 8.97 and 9.88 MeV,
the result is $\sum_{\le 9.88~{\rm MeV}} B_{M2}(E) = 11785~\mu^{2}_{N}$fm$^{2}$.
This value is in agreement with other theoretical predictions and also comparable
with the experimental data in Table \ref{table:Exp_Pb208_02}.
For another example, if we limit our summation up to 8.97 MeV,
then the total strength reduces as $\sum_{\le 8.97~{\rm MeV}} B_{M2}(E) =3694~\mu^{2}_{N}$fm$^{2}$,
which is smaller than the experimental data.
If the higher-energy M2 measurement becomes available, the comparison could improve the agreement.

We again mention the collectivity of M2 and its dependence on mass numbers.
The Weisskopf estimate gives
$B_{SP-M2}\cong 1.45\cdot A^{2/3}=9.21$, $19.2$, and $50.9~\mu^2_{N}{\rm fm}^2$ for
$^{16}$O, $^{48}$Ca, and $^{208}$Pb nuclei, respectively \cite{1951Weisskopf,RingSchuck}.
On the other side, the present RRPA predicts the most excited state in
each nucleus as shown in Tables in the Appendix:
$B_{M2}=487.57~\mu^2_{N}{\rm fm}^2$ at $17.81$ MeV in $^{16}$O;
$B_{M2}=1569.21~\mu^2_{N}{\rm fm}^2$ at $15.08$ MeV in $^{48}$Ca;
$B_{M2}=8020.14~\mu^2_{N}{\rm fm}^2$ at $9.88$ MeV in $^{208}$Pb.
Their ratios thus read $B_{M2}/B_{SP-M2}=52.9$, $81.7$, and $157.4$ for
$^{16}$O, $^{48}$Ca, and $^{208}$Pb nuclei, respectively, indicating the mean
numbers of excited nucleons, that seem to be overestimated. Thus, as reliable measure of collective properties of M2 transitions
we consider our detailed analysis of the $ph$ composition in relevant
excited states, as given in Appendix.

\subsection{Pairing effect in $^{18}$O, $^{42}$Ca, $^{56}$Fe, and $^{90}$Zr} \label{sec:PAIRING}
Of particular interest here are the effects of pairing correlations on the M2 transitions.
Our calculations are based on the DD-PC1 parametrization for the relativistic EDF \cite{Niksic03}, and pairing correlations are described using
the pairing part of the Gogny interaction \cite{Berger}.

\begin{figure}[tb]
\includegraphics[scale=0.33]{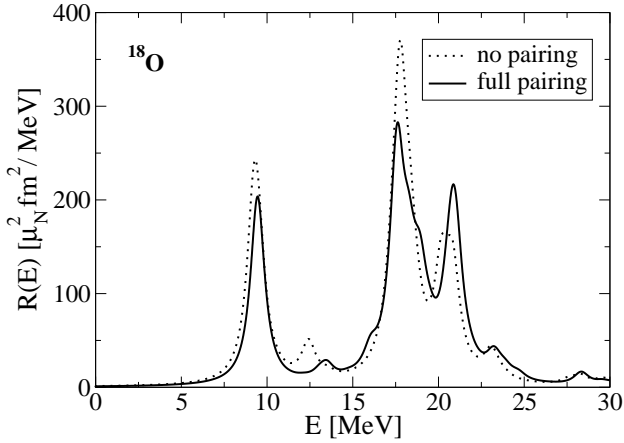}
\caption{The RQRPA full M2 response function for $^{18} \rm O$, $R_{M2}(E)$, based on the DD-PC1 parameterization and Gogny-pairing interaction.
The response functions without pairing correlations and with full pairing in the
residual RQRPA interaction are shown separately.}
\label{fig:O18_01}
\end{figure}

\begin{figure}[th]
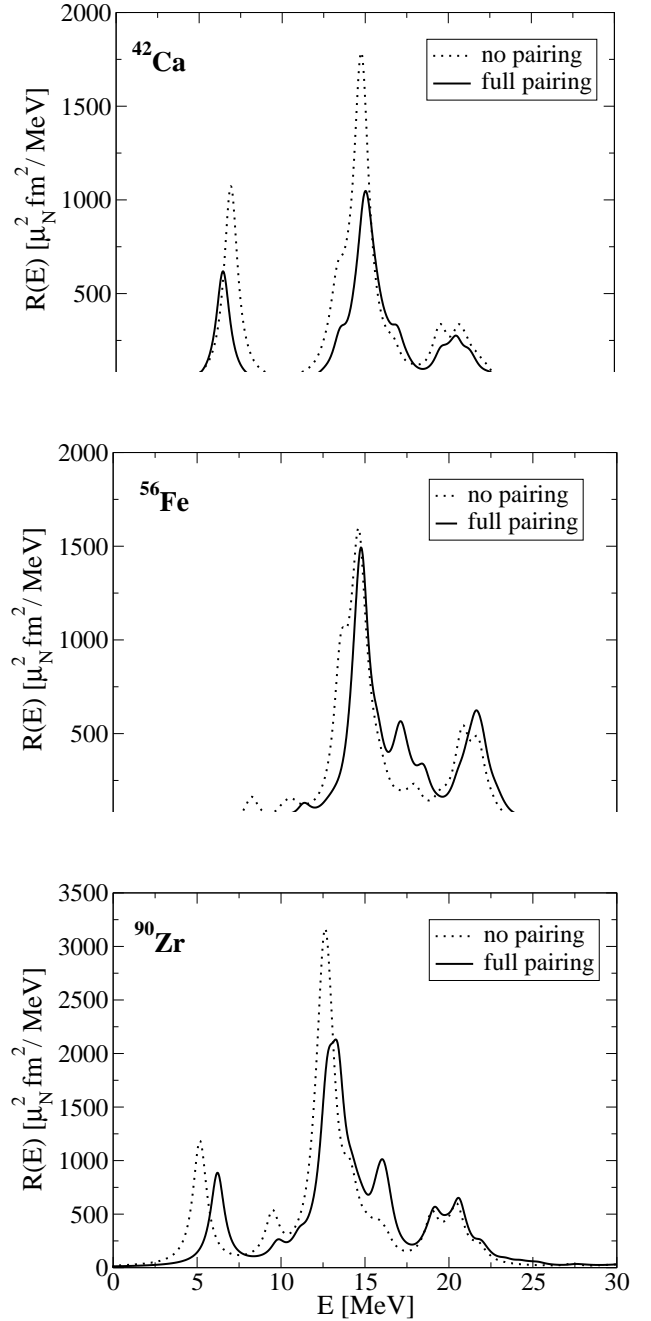

\includegraphics[scale=0.33]{M2_Ca42_PAIRING_MTD4_EPS600.eps}
\includegraphics[scale=0.33]{M2_Fe56_PAIRING_MTD4_EPS600.eps}
\includegraphics[scale=0.33]{M2_Zr90_PAIRING_MTD4_EPS600.eps}
\caption{The same as Fig.~\ref{fig:O18_01} but for $^{42}$Ca, $^{56}$Fe, and $^{90}$Zr nuclei.}
\label{fig:Ca42_01}
\end{figure}


Figure \ref{fig:O18_01} shows the RQRPA results for the full response function for M2 transitions of $^{18}$O.
The response function without pairing correlations in the residual RQRPA interaction is also shown.
In comparison to previously studied nucleus $^{16}$O,
open-shell system $^{18}$O displays rather different response.
In particular, its high-energy part is fragmented into two strong peaks.
The third peak at 21 MeV is indeed from valence neutrons in the $d_{5/2}$ orbit, which was empty in the previous $^{16}$O case.
By comparing two cases with and without pairing correlations,
one can observe considerable pairing effects on the transition strengths,
whereas their excitation energies are less sensitive.
We perform similar analysis for $^{42}$Ca and $^{56}$Fe as shown in Fig. \ref{fig:Ca42_01}.
The response function for $^{42}$Ca displays a triple-hump structure,
and its strength is strongly sensitive on pairing correlations.
For $^{56}$Fe, the low-energy peak is absent in the full-pairing calculation,
and the overall transition strength is rather fragmented.

We next investigate the $^{90}$Zr nucleus, because experimental data on M2 transitions
have been reported up to the excitation energy of about 12 MeV in Ref. \cite{Cosel01}.
The calculated M2 response function for $^{90}$Zr is shown in Fig. \ref{fig:Ca42_01}.
There, three main structures are obtained in the strength distribution,
with the most pronounced peak around 13 MeV.
The RQRPA result for the EW summation amounts $m_{1}(E) = 160.10 \times 10^{3}~{\rm  MeV}\mu^{2}_{N}$fm$^{2}$ up to 50 MeV.
For comparison, the experimental data from Ref. \cite{Cosel01} are
available up to 12 MeV, resulting in the value as
$m^{\rm expt.}_{1}(E) = 23.6 \times 10^{3}~{\rm  MeV}\mu^{2}_{N}$fm$^{2}$ for $^{90}$Zr.
This value is considerably lower than our prediction.
If we limit our theoretical consideration up to 12 MeV, the EW summation gives
$m_{1}(E \leq 12~{\rm MeV}) = 13.67\times 10^{3}~{\rm  MeV}   \mu^{2}_{N}$fm$^{2}$,
which is of the reasonable order with respect to the experimental value.
Since several M2 peaks are predicted above 12 MeV,
future experiments may additionally find higher transitions for complete information.

\begingroup\renewcommand{\arraystretch}{1.5}
\begin{table}[tb]
\begin{center}
\scalebox{1.}{
\begin{tabular}{ c c c c }
\hline
\hline
 & Pairing &$m_0$   &$\bar{E}$  \\
 &         &[$10^{3} \mu^{2}_{N}$fm$^{2}$]  &[MeV]  \\
 \hline  
$^{18}$O  &  no   &  16.17 &  16.87 \\
         &  full  &  15.40    & 17.91\\
\hline
$^{42}$Ca &  no   &  6.89 &  14.06 \\
         &  full  & 4.62  & 14.81  \\  
\hline
$^{56}$Fe &  no   &  6.29 &  16.47   \\
         &  full  &  5.87   & 17.80  \\    
\hline
$^{90}$Zr &  no   &  11.78 &  13.09   \\
         &  full  & 11.02  & 14.53 \\       
\hline
\hline
\end{tabular}%
}
\end{center}
\caption{The NEW summation, $m_0=\sum_{E} B_{M2}(E)$ up to 50 MeV, and the centroid energy, $\bar{E}=m_1/m_0$, calculated with the present RQRPA for $^{18} \rm O$, $^{42}$Ca, $^{56}$Fe, and $^{90}$Zr.
Results with and without pairing interaction are compared.}
\label{table:Th_O18-Zr90}
\end{table}
\renewcommand{\arraystretch}{1}

Another experimental investigation with the inelastic electron
scattering \cite{Muller01}, whose data were analized with the
distorted-wave Born approximation, provides the NEW-M2 summation,
$\sum B^{\rm expt.}_{M2}(E) = 950\pm 100~\mu^{2}_{N}$fm$^{2}$ for
the M2-excited energies between 8-10 MeV for $^{90}$Zr.
For comparison, from our RQRPA, the same summation amounts $291~\mu^{2}_{N}$fm$^{2}$
in the case with pairing correlation.
Indeed in our RQRPA results between 8-10 MeV, there is only one M2-excited peak predicted,
and thus, the NEW summation becomes smaller than the experimental one.

 {The non-relativistic RPA study for $^{90}$Zr exists also in Ref. \cite{Cosel01},
where the EW sum of the B(M2) strength up to 12 MeV 
yields $m_{1}= 112.3\times 10^{3}~{\rm  MeV}\mu^{2}_{N}$fm$^{2}$ 
(including the quenching of 0.6 in the spin $g$ factor),
being higher than the experimental value. 
Similar to the present study,
the second RPA calculation \cite{Cosel01}, that is based on rather
different effective nuclear interactions and it does not include pairing,
also result in pronounced low-energy M2 spin excitation, at $\approx$ 5 MeV. Therefore, one can
expect that by extending presently experimentally covered energy 
range between 7-12 MeV, both to low-energy and high-energy regions, would 
result in more M2 transitions that would reduce presently existing 
discrepancies between the overall measured and calculated strengths.
We note that by including couplings to complex configurations such
as $2p2h$ in the second RPA \cite{Cosel01}, considerable fragmentation
of the transition strength is obtained when compared to calculation
based only at the RPA level. Therefore, it is expected that the 
structure predicted by the RQRPA would become much more
fragmented if complex configurations would be included. 
}

In Table \ref{table:Th_O18-Zr90} we summarize the NEW summations and centroid energies for open-shell nuclei with and without full consideration of the pairing correlation.
In general, the pairing interaction reduces the NEW summation and shifts the centroid energy to the higher region.

\subsection{M2 transitions in $^{36-64} \rm Ca$ isotopes}\label{sec:CaISOTOPES}
In order to assess how the M2 transition evolves with the increase of the
neutron number along the isotope chain, we have calculated the RQRPA
response functions in  $^{36-64} \rm Ca$  isotopes. Calculated
predictions on M2 excitation properties for Ca isotopes could be useful
for the future experimental studies.
Two relativistic density-dependent interactions have been used, DD-PC1 \cite{Niksic03} and DD-PCX \cite{Yuksel2019}, supplemented with the Gogny force \cite{Berger} for the 
pairing correlations.

The calculated M2 response functions are shown for $^{36-64} \rm Ca$ in Fig. \ref{fig:Ca_DDPC1_DDPCX}.
The transition strength distributions show no significant differences when 
comparing DD-PC1 versus DD-PCX results, neither in terms of fragmentation 
or transition strength.
With increasing the neutron number, the overall M2 transition strength increases. 
Qualitative changes are especially pronounced in the low-energy 
part of the spectra which becomes more fragmented, and in neutron-rich 
Ca isotopes the main M2 peak becomes split into two peaks, which are of comparable strength for $^{60,64} \rm Ca$.

\begin{figure}[tb]
\includegraphics[scale=0.45]{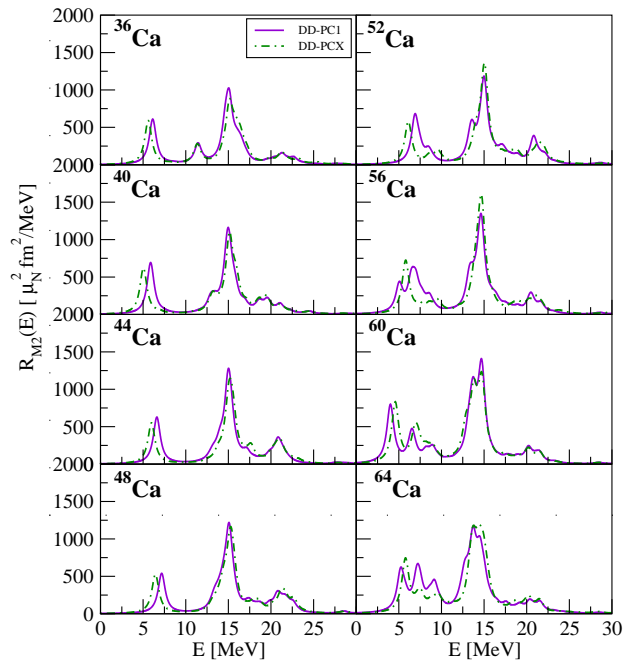}
\caption{The M2 full RQRPA response functions for Ca isotopes,
shown for the DD-PC1 and DD-PCX interactions.}
\label{fig:Ca_DDPC1_DDPCX}
\end{figure}

In Fig. \ref{fig:Ca_ISOTOPES} the M2 response is shown separately 
for the $R^\sigma_{M2}(E)$ (spin) and $R^\ell_{M2}(E)$ (orbital) transitions.
Similarly in the case of M1 transitions for spherical symmetric
nuclei \cite{Kruzic01}, the spin M2 transition is much stronger in
comparison to almost-negligible orbital one.
Thus the evolution of the M2 response in Ca isotopes is governed mainly by changes in the
spin-response, which also determines the increase of the fragmentation
of the spectra when moving toward neutron-rich isotopes. 
Due to small contribution of the orbital-response, it is difficult to conclude on
its evolution along the isotope chain.

\begin{figure}[t]
\includegraphics[scale=0.45]{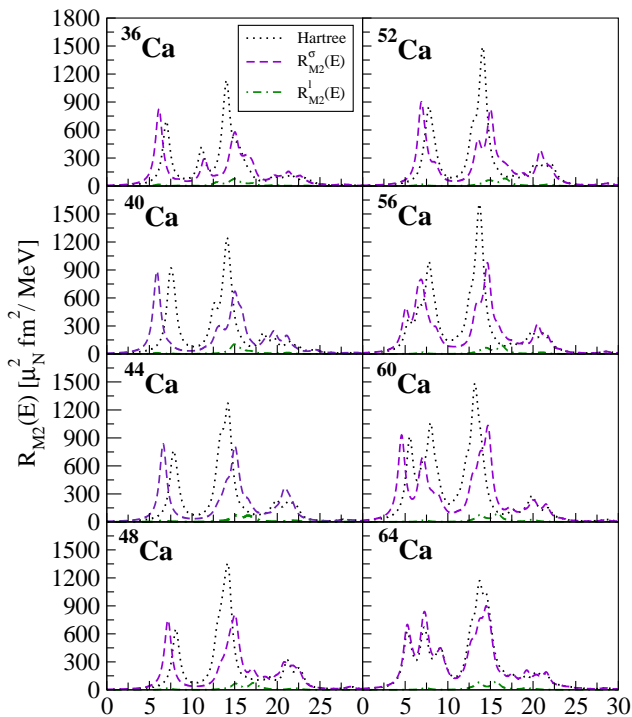}
\caption{The M2 spin and orbital RQRPA response functions supplemented with the unperturbed RHB response for Ca isotopes, using DD-PC1 interaction.}
\label{fig:Ca_ISOTOPES}
\end{figure}

Finally, in Fig. \ref{fig:Ca_m0_m1}, the $m_0$ (NEW) and $m_1$ (EW)
M2 summations are shown for $^{36-64}$Ca isotope chain, using the
RQRPA with the DD-PC1 and DD-PCX interactions.
In this way we can observe the evolution of the M2 properties with
increasing the neutron number in Ca isotopes.
The model calculations are performed using two limit values for the quenching
of the spin and orbital $g$ factors, $\zeta = 0.93$ and $\zeta = 0.8$,
which were previously discussed in the study of M1 transitions \cite{Kruzic01}. 
Therefore, it allows one to infer how these limits in the quenching factors
fit into the future experimental studies of M2 transitions in Ca isotopes.
The figure also demonstrates rather weak model dependence over the whole 
isotope chain when using DD-PC1 and DD-PCX interactions. 
In Fig. \ref{fig:Ca_m0_m1} we have denoted the overlap of the results obtained from the two interactions,
that represents a theoretical prediction of this study for $m_0$ and $m_1$
moments, to be compared with forthcoming experiments. 
In this way, the analysis of M2 transitions, simultaneously with those of M1 transitions, can provide
deeper insight into the quenching of gyromagnetic factors.

\begin{figure}[t]
\includegraphics[scale=0.33]{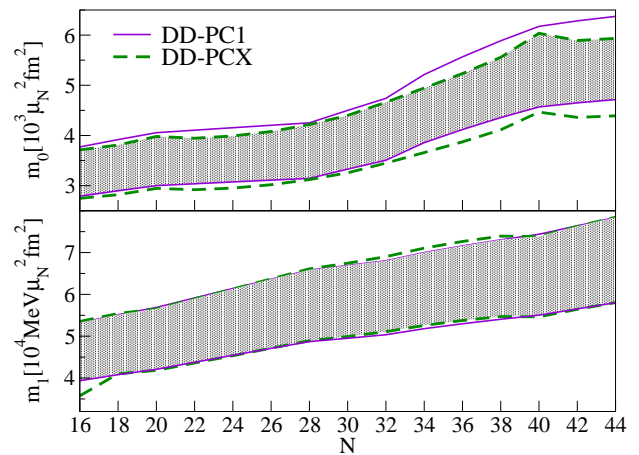}
\caption{The evolution of the $m_0$ and $m_1$ RQRPA moments for
M2 transitions in $^{36-64} \rm Ca$ isotope chain, for DD-PC1 and DD-PCX interactions.
Higher limit values correspond to the quenching of $g$ factors 
$\zeta = 0.93$, and lower limit values to $\zeta = 0.8$.}
\label{fig:Ca_m0_m1}
\end{figure}

\section{Summary}\label{sec:SUM}
In this work we have investigated the properties of M2 transitions
from $0^{+}$ ground state to $2^{-}$ excited states in even-even nuclei, by
using the RHB+RQRPA framework. Two density-dependent relativistic point coupling
interactions have been used, DD-PC1 and DD-PCX, supplemented with the IV-PV
term in the residual RQRPA interaction. The pairing correlations have been 
described with the pairing part of the Gogny force. Model calculations of M2
transitions in $^{16} \rm O$, $^{48} \rm Ca$, and $^{208} \rm Pb$ provided
insight into their structure and collective properties.
The analysis of relevant particle-hole configurations confirmed that
the main M2 peaks have rather complicated structure, different
than single-particle transition, where nucleons in different orbits simultaneously contribute. With increase of the mass number the
collectivity of main M2 states increases due to larger number
of possible $ph$ transitions. The calculated M2 transition strengths
appear larger than available experimental data, possibly because (i) measurements are
available only within rather restricted energy ranges,
and/or (ii) beyond-QRPA effects are missing in the present approach.
Similar results are obtained in earlier RPA and shell-model calculations, 
if no additional quenching is imposed on the $g$ factors.
For the missing low-lying M2 peak in
measurements for $^{48}$Ca, we point out that further
experimental studies of low-energy M2 transitions could provide 
additional constraints for the IV-PV interaction,
and clarify if additional M2 states exist at energies lower
than presently available from the experiment.
More complete experimental studies are expected as reference
for benchmarking the model calculations.
We especially emphasize the interest in two cases, i.e.,
(i) the low-lying region of $^{48}$Ca, where another peak of proton's
$1d^{-1}_{3/2}\longrightarrow 1f_{7/2}$ transition is predicted, and
(ii) the region higher than 9 MeV of $^{208}$Pb, where its M2 response
is predicted to have rather distributed structure.

We have also demonstrated that pairing effects 
in open-shell nuclei $^{18} \rm O$, $^{42}$Ca, $^{56}$Fe, and semi-magic $^{90}$Zr
have an impact on M2 transition spectra, by reducing the strength and 
shifting the centroid energy to the higher values.
 {The evolution of the $m_0$ and $m_1$ moments in $^{36-64}$Ca 
isotope chain has also been investigated.}
By imposing the limit values for the quenching of the $g$ factors consistent with the analysis of M1 transitions \cite{Kruzic01},
we provide the limits for $m_0$ and $m_1$ moments for M2 transitions 
within the Ca isotopes, that could provide useful constraints
for the quenching of $g$ factors in nuclear medium.
To our present knowledge, there in no systematic experimental data on properties
of M2 transitions available for Ca isotopes except $\rm ^{48}Ca$.
Thus, we hope that the predictions of this work could motivate future experimental studies on M2 transitions in Ca isotopes.
Some preparations along this line of experimental
research are already in progress \cite{pvc01}.

\section*{acknowledgements}
We thank Peter von Neumann-Cosel for useful discussion on magnetic transitions.
This work is supported within the Tenure Track Pilot Programme of the Croatian Science Foundation and the \'{E}cole Polytechnique F\'{e}d\'{e}rale de Lausanne, and the Project TTP-2018-07-3554 Exotic Nuclear Structure and Dynamics, with funds of the Croatian-Swiss Research Programme.
This work is supported by the ``QuantiXLie Centre of Excellence'' project co-financed by the Croatian Government and European Union through the European Regional Development Fund, the Competitiveness and Cohesion Operational Programme (code KK.01.1.1.01.0004).
We acknowledge support by the Multidisciplinary Cooperative Research Program of the Center for Computational Sciences, University of Tsukuba, using Oakforest-PACS Systems (project No. xg21i064, FY2021), and T.O. acknowledges the support by Takashi Nakatsukasa and Hiroyuki Kobayashi in this program.

\appendix
\section{The analysis of contributing particle-hole transitions in M2 excited states}
Here we display the components of M2 transitions for $^{16}$O, $^{48}$Ca, and $^{208}$Pb nuclei, on which discussions are presented in the main text.

\newpage
\renewcommand{\arraystretch}{1.5}
\begin{table}[t]
\begin{center}
\begin{tabular}{c c c c c}
\hline
\hline
$E^{th.}_{peak}$    &$B_{M2}(E)$                      &$b^{\nu}_{ph}$ &$b^{\pi}_{ph}$ &$ph$ transition  \\ 
$\lbrack \rm MeV \rbrack$ &$\lbrack \mu^{2}_{N} \rm fm^{2} \rbrack$  &$\lbrack \mu_{N} \rm fm \rbrack$ &$\lbrack \mu_{N}\rm  fm \rbrack$ &   \\ 
 \hline 
 8.60      & 397.16   &  -8.43  &   -13.72  &  $(\nu 1p^{-1}_{1/2}  \rightarrow 1d_{5/2})$ \\
      &       &     &        &  $(\pi  1p^{-1}_{1/2}   \rightarrow 1d_{5/2} )$ \\
      &       & -0.30   &       2.85  &  $(\nu 1p^{-1}_{3/2}   \rightarrow 2s_{1/2} )$ \\
      &       &     &        &  $(\pi  1p^{-1}_{3/2}   \rightarrow 1d_{5/2} )$ \\
\hline
 17.15    & 206.73  & 1.85 &  -1.29      &  $(\nu 1p^{-1}_{1/2}  \rightarrow 1d_{3/2})$ \\
      &      &  &          &  $(\pi 1p^{-1}_{1/2}   \rightarrow 1d_{5/2} )$ \\
      &      & 0.87  &   -15.81    &  $(\nu 1p^{-1}_{3/2}   \rightarrow 1d_{5/2} )$ \\
      &      &  &          &  $(\pi 1p^{-1}_{3/2}   \rightarrow 1d_{5/2} )$ \\
\hline
 17.81  & 487.57 &-1.35  &  13.99  & $(\nu 1p^{-1}_{3/2}  \rightarrow 2s_{1/2})$ \\
    &        &  &     & $(\pi 1p^{-1}_{3/2}   \rightarrow  1d_{5/2} )$ \\
    &        &1.46  &  0.35    & $(\nu 1p^{-1}_{1/2}  \rightarrow  1d_{3/2})$ \\
    &        &  &     & $(\pi 1p^{-1}_{1/2}   \rightarrow  1d_{5/2} )$ \\
    &        &1.48  &  -0.23   & $(\nu 1p^{-1}_{1/2}  \rightarrow 1d_{5/2})$ \\
    &        & &     & $(\pi 1p^{-1}_{3/2}   \rightarrow  2s_{1/2} )$ \\
    &        &6.96  &\text{0} & $(\nu 1p^{-1}_{3/2}  \rightarrow 1d_{5/2})$ \\
\hline 
 19.04    & 158.74   &-2.53  & -11.93  & $(\nu 1p^{-1}_{3/2}  \rightarrow 1d_{5/2})$ \\
      &        &  &      & $(\pi 1p^{-1}_{3/2}   \rightarrow  1s_{1/2} )$ \\
\hline
 19.99   & 123.27   &-8.47  &  -2.29  & $(\nu 1p^{-1}_{3/2}  \rightarrow 2s_{1/2})$ \\
      &        &  &     & $(\pi 1p^{-1}_{3/2}   \rightarrow 1d_{5/2} )$ \\
\hline
\hline
\end{tabular}
\end{center}
\caption{Partial contributions $b^{\nu}_{ph}$ (neutrons) and  $b^{\pi}_{ph}$ (protons) to the dominant M2 transitions in the $^{16}$O nucleus.
The $E^{th.}_{peak}$ denotes the peak energy corresponding to the RRPA eigenvalue, and $B_{M2}(E)$ is its overall transition strength.}
\label{table:Th_O16_01}
\end{table}
\renewcommand{\arraystretch}{1}

\renewcommand{\arraystretch}{1.5}
\begin{table}[t]
\begin{center}
\begin{tabular}{ c c c c c c c}
\hline
\hline
$E^{th.}_{peak}$    &$B_{M2}(E)$                      &$b^{\nu}_{ph}$ &$b^{\pi}_{ph}$ & $ph$ transition  \\ 
$\lbrack \rm MeV \rbrack$ &$\lbrack \mu^{2}_{N} \rm fm^{2} \rbrack$  &$\lbrack \mu_{N} \rm fm \rbrack$ &$\lbrack \mu_{N}\rm  fm \rbrack$ &   \\ 
 \hline
 7.17    & 834.41 & -0.76  &   -29.69  &  $(\nu 2s^{-1}_{1/2}  \rightarrow 2p_{3/2})$ \\
   &      &   &      &  $(\pi 1d^{-1}_{3/2}   \rightarrow 1f_{7/2} )$ \\
   &      & -0.59  &      3.56  &  $(\nu 1f^{-1}_{7/2}   \rightarrow 1g_{9/2} )$ \\
   &      &   &      &  $(\pi 1d^{-1}_{5/2}   \rightarrow 1f_{7/2} )$ \\
\hline
 13.28  & 122.26  &  -6.85  &   -0.13    &  $(\nu 1f^{-1}_{7/2}  \rightarrow 2d_{3/2})$ \\
   &      &    &       &  $(\pi 1d^{-1}_{5/2} \rightarrow 2p_{1/2} )$ \\
   &      &    1.73  &     -0.18  &  $(\nu 2s^{-1}_{1/2} \rightarrow 2p_{3/2} )$ \\
   &      &    &       &  $(\pi 1d^{-1}_{3/2}   \rightarrow 2p_{3/2} )$ \\
   &      &  -2.02   &     -0.18  &  $(\nu 1d^{-1}_{3/2}   \rightarrow 2p_{3/2} )$ \\
   &      &    &       &  $(\pi 1d^{-1}_{5/2}   \rightarrow 2p_{3/2} )$ \\
   &      &  -1.12   &    -0.19   &  $(\nu 1f^{-1}_{7/2}   \rightarrow 2d_{5/2} )$ \\
   &      &    &      &  $(\pi 1d^{-1}_{3/2}   \rightarrow 1f_{7/2} )$ \\
   &      &   -2.36  & \text{0} &  $(\nu 1f^{-1}_{7/2}   \rightarrow 1g_{9/2} )$ \\
\hline
 14.00  & 180.43 &  -13.73 &   7.85    &  $(\nu 2s^{-1}_{1/2}  \rightarrow 2p_{3/2})$ \\
   &      &   &      &  $(\pi  2s^{-1}_{1/2}   \rightarrow 2p_{3/2} )$ \\
   &      &   -7.69 &   1.00    &  $(\nu  1f^{-1}_{7/2}   \rightarrow 1g_{9/2} )$ \\
   &      &   &      &  $(\pi 1d^{-1}_{3/2}   \rightarrow 1f_{7/2} )$ \\
\hline
 14.41  & 223.95 &  -5.05  &   1.47     &  $(\nu 2s^{-1}_{1/2}  \rightarrow 2p_{3/2})$ \\
   &      &   &           &  $(\pi 2s^{-1}_{1/2}   \rightarrow 2p_{3/2} )$ \\
   &      &   -1.73 &   8.68      &  $(\nu  1f^{-1}_{7/2}   \rightarrow 1g_{9/2} )$ \\
   &      &   &          &  $(\pi   1d^{-1}_{5/2}   \rightarrow 1f_{7/2} )$ \\
\hline
 14.61  & 159.10 &  -1.11  &   -9.32   &  $(\nu 2s^{-1}_{1/2}  \rightarrow 2p_{3/2})$ \\
    &      &   &      &  $(\pi 2s^{-1}_{1/2}   \rightarrow 2p_{3/2} )$ \\
    &      &  -5.59  &   -1.07   &  $(\nu  1f^{-1}_{7/2}   \rightarrow 1g_{9/2} )$ \\
    &      &   &      &  $(\pi 1d^{-1}_{3/2}   \rightarrow 2p_{3/2} )$ \\
    &      & 0.21   &    1.57    &  $(\nu  1f^{-1}_{7/2}   \rightarrow 2d_{3/2} )$ \\
    &      &   &      &  $(\pi 1d^{-1}_{3/2}   \rightarrow 1f_{7/2} )$ \\
    &      &  0.27  &    2.29   &  $(\nu 1d^{-1}_{3/2}   \rightarrow 2p_{3/2} )$ \\
    &      &   &      &  $(\pi 1d^{-1}_{5/2}   \rightarrow 1f_{7/2} )$ \\
\hline
\hline 
\end{tabular}%
\end{center}
\caption{The same as Table~\ref{table:Th_O16_01} but for $^{48} \rm Ca$.}
\label{table:Th_Ca48_01}
\end{table}
\renewcommand{\arraystretch}{1}

\renewcommand{\arraystretch}{1.5}
\begin{table}[t]
\begin{center}
\begin{tabular}{ c c c c c c c}
\hline
\hline
$E^{th.}_{peak}$    &$B_{M2}(E)$                      &$b^{\nu}_{ph}$ &$b^{\pi}_{ph}$ & $ph$ transition  \\ 
$\lbrack \rm MeV \rbrack$ &$\lbrack \mu^{2}_{N} \rm fm^{2} \rbrack$  &$\lbrack \mu_{N} \rm fm \rbrack$ &$\lbrack \mu_{N}\rm  fm \rbrack$ &   \\ 
 \hline
 15.08  & 1569.21  & 1.44 &  -6.82&  $(\nu 1d^{-1}_{3/2}  \rightarrow 1f_{5/2})$ \\
    &     & &         &  $(\pi 2s^{-1}_{1/2}   \rightarrow 2p_{3/2} )$ \\
    &     & 0.61  &   -2.17      &  $(\nu 1d^{-1}_{3/2}   \rightarrow 2p_{1/2} )$ \\
    &     & &         &  $(\pi 1d^{-1}_{3/2}   \rightarrow 1f_{7/2} )$ \\
    &     & -0.69 &  -33.23     &  $(\nu 2s^{-1}_{1/2}   \rightarrow 2p_{3/2} )$ \\
    &     & &         &  $(\pi 1d^{-1}_{5/2}   \rightarrow 1f_{7/2} )$ \\
\hline
 15.57    & 192.15    &  1.61   &   -3.11    &  $(\nu 1d^{-1}_{3/2}  \rightarrow 2p_{1/2})$ \\
      &      &   &      &  $(\pi 2s^{-1}_{1/2}   \rightarrow 2p_{3/2} )$ \\
      &      &  2.73   &   13.41   &  $(\nu 2s^{-1}_{1/2}   \rightarrow 2p_{3/2} )$ \\
      &      &   &      &  $(\pi 1d^{-1}_{5/2}   \rightarrow 1f_{7/2} )$ \\
      &      & -1.49   &   0.94    &  $(\nu 1d^{-1}_{3/2}   \rightarrow 1f_{7/2} )$ \\
      &      &   &      &  $(\pi 1d^{-1}_{3/2}   \rightarrow 1f_{7/2} )$ \\
\hline
 19.81    & 119.60 &  2.83   &   -1.22    &  $(\nu 1d^{-1}_{5/2}  \rightarrow 2p_{1/2})$ \\
     &      &   &      &  $(\pi 1d^{-1}_{5/2}   \rightarrow 2p_{3/2} )$ \\
     &      &  -1.29  &  -1.19    &  $(\nu 1f^{-1}_{7/2}     \rightarrow 3d_{3/2} )$ \\
     &      &   &      &  $(\pi 1d^{-1}_{5/2}    \rightarrow 1f_{7/2} )$ \\
     &      &   -8.32 &  -0.48    &  $(\nu 1d^{-1}_{5/2}   \rightarrow 2p_{3/2} )$ \\
     &      &   &      &  $(\pi 2s^{-1}_{1/2}    \rightarrow 2p_{3/2} )$ \\
\hline
 20.74   & 302.61     & -9.01  &  -1.22   & $(\nu 1d^{-1}_{5/2}  \rightarrow 2p_{1/2})$ \\
    &        &  &     & $(\pi 1d^{-1}_{5/2}   \rightarrow  2p_{1/2} )$ \\
    &        & -2.38  &  -1.46   & $(\nu 1d^{-1}_{5/2}  \rightarrow 2p_{3/2})$ \\
    &        &  &     & $(\pi 1d^{-1}_{5/2}   \rightarrow  1f_{7/2} )$ \\
    &        &  -2.32 &  -0.74   & $(\nu 1f^{-1}_{7/2}   \rightarrow  2g_{7/2} )$ \\
    &        &  &     & $(\pi 1d^{-1}_{5/2}   \rightarrow  1f_{5/2} )$ \\
\hline
 21.59   & 165.43 &  -1.49  &   2.76     &  $(\nu 1f^{-1}_{7/2}  \rightarrow 2g_{7/2})$ \\
    &      &   &       &  $(\pi 1d^{-1}_{5/2}   \rightarrow 2p_{1/2} )$ \\
    &      &  -0.63 &    -14.92 &  $(\nu 1d^{-1}_{5/2}   \rightarrow 2p_{1/2} )$ \\
    &      &   &       &  $(\pi 1d^{-1}_{5/2}   \rightarrow 2p_{3/2} )$ \\
\hline
\hline 
\end{tabular}
\end{center}
\caption{Continuation of Table~\ref{table:Th_Ca48_01} for $^{48} \rm Ca$.}
\label{table:Th_Ca48_03}
\end{table}
\renewcommand{\arraystretch}{1}

\newpage
\renewcommand{\arraystretch}{1.5}
\begin{table}[t]
\begin{center}
\begin{tabular}{ c c c c c c}
\hline
\hline
$E^{th.}_{peak}$    &$B_{M2}(E)$                      &$b^{\nu}_{ph}$ &$b^{\pi}_{ph}$ & $ph$ transition  \\ 
$\lbrack \rm MeV \rbrack$ &$\lbrack \mu^{2}_{N} \rm fm^{2} \rbrack$  &$\lbrack \mu_{N} \rm fm \rbrack$ &$\lbrack \mu_{N}\rm  fm \rbrack$ &   \\
 \hline
 6.46    &  1815.74 &  1.86  & 7.67  & $(\nu 3p^{-1}_{1/2}  \rightarrow 3d_{5/2})$ \\
   &       &  & & $(\pi 2d^{-1}_{3/2} \rightarrow 2f_{7/2} )$ \\ 
   &       & 39.87 & -3.21 & $(\nu 2f^{-1}_{5/2}  \rightarrow 2g_{9/2})$ \\
   &       &  & & $(\pi 2d^{-1}_{5/2} \rightarrow 2f_{7/2} )$ \\ 
   &       & -1.88 & -3.35 & $(\nu 2f^{-1}_{7/2}  \rightarrow 2g_{9/2})$ \\
   &       &  & & $(\pi 1h^{-1}_{11/2} \rightarrow 1i_{13/2} )$ \\ 
   &       & -1.11 &-0.48  & $(\nu 1i^{-1}_{13/2}  \rightarrow 1j_{15/2})$ \\
   &       &  & & $(\pi 3s^{-1}_{1/2} \rightarrow 3p_{3/2} )$ \\ 
\hline 
 7.45    &  982.58& -23.71 & -18.55  & $(\nu 3p^{-1}_{1/2}  \rightarrow 3d_{5/2})$ \\
   &      &   &    & $(\pi 2d^{-1}_{3/2} \rightarrow 2f_{7/2} )$ \\
   &      &   1.37  & 2.61     & $(\nu 3p^{-1}_{3/2}   \rightarrow 3d_{5/2} )$ \\
   &      &   &    & $(\pi 2d^{-1}_{5/2} \rightarrow 2f_{7/2} )$ \\
   &      &  7.82  & 1.65      & $(\nu 2f^{-1}_{5/2}   \rightarrow 2g_{9/2} )$ \\
   &      &   &    & $(\pi  1h^{-1}_{11/2} \rightarrow 1i_{13/2} )$ \\
\hline 
 7.61   &  420.15 &-12.38 & 1.22       & $(\nu 3p^{-1}_{1/2}  \rightarrow 3d_{5/2})$ \\
   &       &   &       & $(\pi   3s^{-1}_{3/2} \rightarrow 3p_{3/2} )$ \\
   &       &   2.51 & 37.45    & $(\nu 3p^{-1}_{3/2}  \rightarrow 3d_{5/2})$ \\
   &       &   &       & $(\pi 2d^{-1}_{3/2} \rightarrow 2f_{7/2} )$ \\
   &       & -7.23 & -4.07     & $(\nu   2f^{-1}_{5/2}  \rightarrow 2g_{9/2})$ \\
   &       &   &       & $(\pi 2d^{-1}_{5/2} \rightarrow 2f_{7/2} )$ \\
\hline
 8.69   &  336.46 & 1.09   & -3.02      & $(\nu 3p^{-1}_{3/2}  \rightarrow 4s_{1/2})$ \\
   &       &   &       & $(\pi   2d^{-1}_{3/2} \rightarrow 2f_{7/2} )$ \\
   &       & 2.49   & -2.09     & $(\nu 3p^{-1}_{1/2}  \rightarrow 3d_{5/2})$ \\
   &       &   &       & $(\pi     1h^{-1}_{11/2} \rightarrow 1i_{13/2} )$ \\
   &       & 24.13 & -0.34     & $(\nu 3p^{-1}_{3/2}  \rightarrow 3d_{5/2})$ \\
   &       &   &       & $(\pi   2d^{-1}_{5/2} \rightarrow 2f_{7/2} )$ \\
\hline
 8.97   &  231.85 & 16.78 & -1.03      & $(\nu 3p^{-1}_{3/2}  \rightarrow 4s_{1/2})$ \\
   &       &   &       & $(\pi   2d^{-1}_{5/2} \rightarrow 2f_{7/2} )$ \\
   &       & -1.62 & -1.69      & $(\nu  3p^{-1}_{3/2}  \rightarrow 3d_{5/2})$ \\
   &       &   &       & $(\pi     1h^{-1}_{11/2} \rightarrow 1i_{13/2} )$ \\
   &       & -1.24 & 0.37       & $(\nu    2f^{-1}_{7/2}  \rightarrow 2g_{9/2})$ \\
   &       &   &       & $(\pi   2d^{-1}_{5/2} \rightarrow 3p_{3/2} )$ \\
\hline
\hline
\end{tabular}%
\end{center}
\caption{The same as Table~\ref{table:Th_O16_01} but for $^{208} \rm Pb$.}
\label{table:Th_Pb208_01}
\end{table}
\renewcommand{\arraystretch}{1}

\renewcommand{\arraystretch}{1.5}
\begin{table}[t]
\begin{center}
\begin{tabular}{ c c c c c c}
\hline
\hline
$E^{th.}_{peak}$    &$B_{M2}(E)$                      &$b^{\nu}_{ph}$ &$b^{\pi}_{ph}$ & $ph$ transition  \\ 
$\lbrack \rm MeV \rbrack$ &$\lbrack \mu^{2}_{N} \rm fm^{2} \rbrack$  &$\lbrack \mu_{N} \rm fm \rbrack$ &$\lbrack \mu_{N}\rm  fm \rbrack$ &   \\ 
\hline 
 9.88    &  8020.14 &  1.58 & -1.62    & $(\nu 2f^{-1}_{5/2}  \rightarrow 1g_{9/2})$ \\
   &      &  &    & $(\pi 3s^{-1}_{1/2} \rightarrow 1p_{3/2} )$ \\
   &      &   9.84 & -1.16   & $(\nu 2f^{-1}_{7/2}   \rightarrow 1g_{9/2} )$ \\
   &      &  &    & $(\pi 2d^{-1}_{3/2} \rightarrow 1f_{7/2} )$ \\
   &      &  -2.19 &-20.72  & $(\nu 1h^{-1}_{9/2}   \rightarrow 1i_{11/2} )$ \\
   &      &  &    & $(\pi 2d^{-1}_{5/2} \rightarrow 1f_{7/2} )$ \\
   &      & 11.55 &  92.71  & $(\nu 1i^{-1}_{13/2}   \rightarrow 1j_{15/2} )$ \\
   &      &  &    & $(\pi 1h^{-1}_{11/2}  \rightarrow 1i_{13/2} )$ \\
\hline
10.18  &  7843.68 &  -3.91  & 4.62    &  $(\nu 2f^{-1}_{5/2}   \rightarrow 1g_{7/2})$ \\
   &       &    &   & $(\pi 3s^{-1}_{1/2}   \rightarrow 1p_{3/2} )$ \\
   &       &   -1.62 & -3.22   & $(\nu 2f^{-1}_{5/2}   \rightarrow 1g_{9/2} )$ \\
   &       &    &   & $(\pi 2d^{-1}_{3/2}   \rightarrow 1f_{7/2} )$ \\
   &       &   2.71   &-46.81 & $(\nu 1h^{-1}_{9/2}   \rightarrow 1i_{11/2} )$ \\
   &       &    &   & $(\pi 2d^{-1}_{5/2}   \rightarrow 1f_{7/2} )$ \\
   &       &  -13.08 &-27.37 & $(\nu 1i^{-1}_{13/2}   \rightarrow 1j_{15/2} )$ \\
   &       &    &   & $(\pi 1h^{-1}_{11/2}   \rightarrow 1i_{13/2} )$ \\
\hline
 10.78 & 1926.70 &  1.18  & 32.23  &  $(\nu 2f^{-1}_{5/2}   \rightarrow 2g_{9/2})$ \\
  &     &   &  & $(\pi 3s^{-1}_{1/2}   \rightarrow 3p_{3/2} )$ \\
  &     &  -1.24  &  4.75 & $(\nu 1h^{-1}_{9/2}   \rightarrow 1i_{11/2} )$ \\
  &     &   &  & $(\pi 2d^{-1}_{5/2}   \rightarrow 2f_{7/2} )$ \\
  &     &   2.23  & 4.14  & $(\nu 1i^{-1}_{13/2}   \rightarrow 1j_{15/2} )$ \\
  &     &   &  & $(\pi 1h^{-1}_{11/2}   \rightarrow 1i_{13/2} )$ \\
\hline 
 16.24  & 1094.44 &  2.72    & -4.61&  $(\nu 2f^{-1}_{7/2}   \rightarrow 4d_{3/2})$ \\
   &     &     &        & $(\pi 1g^{-1}_{9/2}   \rightarrow 1h_{9/2} )$ \\
   &     &  13.69   &   0.68       & $(\nu 1h^{-1}_{11/2}   \rightarrow 1i_{11/2} )$ \\
   &     &     &        & $(\pi 1g^{-1}_{9/2}   \rightarrow 2f_{5/2} )$ \\
   &     &  6.30     &  15.49      & $(\nu 1i^{-1}_{13/2}   \rightarrow 1j_{13/2} )$ \\
   &     &     &        & $(\pi 1h^{-1}_{11/2}   \rightarrow 1i_{11/2} )$ \\
\hline 
 19.33  & 1888.20 &  16.05  & 5.45  &  $(\nu 1h^{-1}_{11/2}   \rightarrow 2g_{7/2})$ \\
  &      &     & & $(\pi 1h^{-1}_{11/2}   \rightarrow 2g_{7/2} )$ \\
  &      &  8.58     & 5.01   & $(\nu 1i^{-1}_{13/2}   \rightarrow 3h_{9/2} )$ \\
  &      &     & & $(\pi 1g^{-1}_{9/2}   \rightarrow 2f_{5/2} )$ \\
\hline
\hline 
\end{tabular}
\end{center}
\caption{Continuation of Table~\ref{table:Th_Pb208_01} for $^{208} \rm Pb$.}
\label{table:Th_Pb208_02}
\end{table}

\bibliographystyle{apsrev4-1}
\bibliography{main}

\end{document}